# Title: Present Y chromosomes support the Persian ancestry of Sayyid Ajjal Shams al-Din Omar and Eminent Navigator Zheng He

Authors: Chuan-Chao Wang[1], Ling-Xiang Wang[1], Manfei Zhang[1], Dali Yao[1], Li Jin[1,2,3], Hui Li[1]

**Affiliations**

1. State Key Laboratory of Genetic Engineering and MOE Key Laboratory of Contemporary Anthropology, School of Life Sciences, Fudan University, Shanghai, China.
2. CAS-MPG Partner Institute for Computational Biology, SIBS, CAS, Shanghai, China
3. Institute of Health Sciences, China Medical City, Taizhou, Jiangsu, China
   * Correspondence to: lihui.fudan@gmail.com

**Abstract**
Sayyid Ajjal is the ancestor of many Muslims in areas all across China. And one of his descendants is the famous Navigator of Ming Dynasty, Zheng He, who led the largest armada in the world of 15$^{th}$ century. The origin of Sayyid Ajjal's family remains unclear although many studies have been done on this topic of Muslim history. In this paper, we studied the Y chromosomes of his present descendants, and found they all have haplogroup L1a-M76, proving a southern Persian origin.

Most people now inherit surnames from their fathers; and similarly, most men also inherit the Y chromosomes from their fathers. Therefore, men sharing the same surname are expected to have similar Y chromosomes. When combined with surnames and genealogies, Y chromosome can be used to trace the ancestry of eminent persons in history by studying their present descendants. Successful cases include the inference of Y chromosome haplotype of Emperor CAO Cao[1, 2]. Here, we continue to use this method to provide clues about the genetic ancestry of Sayyid Ajjal Shams al-Din Omar and the eminent navigator Zheng He.

Sayyid Ajjal Shams al-Din Omar is a very famous person in East Asia and the Muslim world. Sayyid Ajjal was a Muslim Khwarezmian in Bukhara before his family surrendered to Genghis Khan. Then, he was appointed as the first provincial governor of Yunnan in history by Yuan Dynasty. Sayyid Ajjal also made great contributions to the widespread presence of Islam in China[3]. Many Muslims across China have taken Sayyid Ajjal as their ancestor, especially the famous explorer and diplomat Zheng He in Ming Dynasty and nowadays the Muslims with surname Na in Yunnan province[4]. Na people are suggested by historical records and their genealogies to be the descendants of Nasr al-Din, the first son of Sayyid Ajjal. Zheng He whose original surname was Ma has also been suggested to be the descendant of Nasr al-Din. However, the ancestry of Sayyid Ajjal is still in debate. He was born in Bukhara (nowadays in Uzbekistan)[3]; however, his ancestors were suggested to have moved around West Asia, North Africa, Europe, and Persia before their final settlement in Central Asia[5]. Some other genealogies even pointed to that the ancestors of Sayyid Ajjal had already arrived in China in the Song Dynasty[6]. Therefore, whether Sayyid Ajjal has a Persian ancestry or not is yet to be validated.

Jianbo Na[7] has tried to discuss this issue in his Master degree thesis. He collected 40 male samples with surname Na from Datong County, Yunnan province and typed 9 Y chromosomal short tandem repeats (STRs). He found that 35 individuals shared the same haplotype (DYS19, 14; DYS389I, 12; DYS389b, 16; DYS390, 22; DYS391, 11; DYS392, 14; DYS393, 11) and two individuals only had one step mutation compared with the above haplotype. However, he didn't type the Y chromosomal single nucleotide polymorphisms (SNPs) and couldn't give a deterministic conclusion about the ancestry of Sayyid Ajjal. Here, we recollected male samples with surname Na and Ma according to the genealogies of Sayyid Ajjal and Zheng He from Datong and Kunming, Yunnan province. We typed Y chromosome SNPs as listed in the latest Y-chromosome phylogenetic tree[8, 9] and 17 Y-Filer STRs. One of the three Na samples shared the same haplotype as mentioned above (DYS19, 14; DYS389I, 12; DYS389b, 16; DYS390, 22; DYS391, 11; DYS392, 14; DYS393, 11; DYS437, 15; DYS438, 11; DYS439, 12; DYS448, 19; DYS456, 17; DYS458, 16; DYS635, 24; H4, 12; DYS385a, 13; DYS385b, 17) and the Ma sample also shared the similar haplotype with only one step mutation at DYS19 for the 17 Y-STRs. The haplogroup of Na and Ma samples were assigned as L1a-M76.

Haplogroup L1a-M76 is found mainly in Eastern Iran, Southern Pakistan, and India. A low frequency of L1a-M76 has also been detected in Saudi Arabia, Nepal, and Central Asia (Fig. 1a)[10-14]. To discern the detailed relationships among the Na and Ma people and other related populations, we constructed a median-joining network based on Y-STR haplotypes within the L-M11 haplogroup (Fig.1b). Many Dravidian (southern India) and Malaysian Indian samples together with four Pakistan (three Balochi and one Makrani), four Afghanistan (one Bagram, one Arab, one Balush, and one Uzbek) samples, and one Na individual formed the root clade of haplogroup L1a-M76. Most of Na samples had one step mutation with the root haplotype and they also clustered a middle size clade with few Malaysian Indian samples. The nearly exclusive clade of Na samples indicates the severe founder effect and the subsequent clan expansion.

The Na and Ma Muslims with the Y chromosome haplogroup L1a-M76 might trace their origin to western South Asia, most of where were occupied by Persia for a long time during the period of their ancient ancestors. The strict Islamic genealogies also link the Na and Ma Muslims to Sayyid Ajjal and Zheng He hundreds of years ago. Thus, the suggested Persian ancestry for Sayyid Ajjal and Zheng He is supported by genetic evidence.

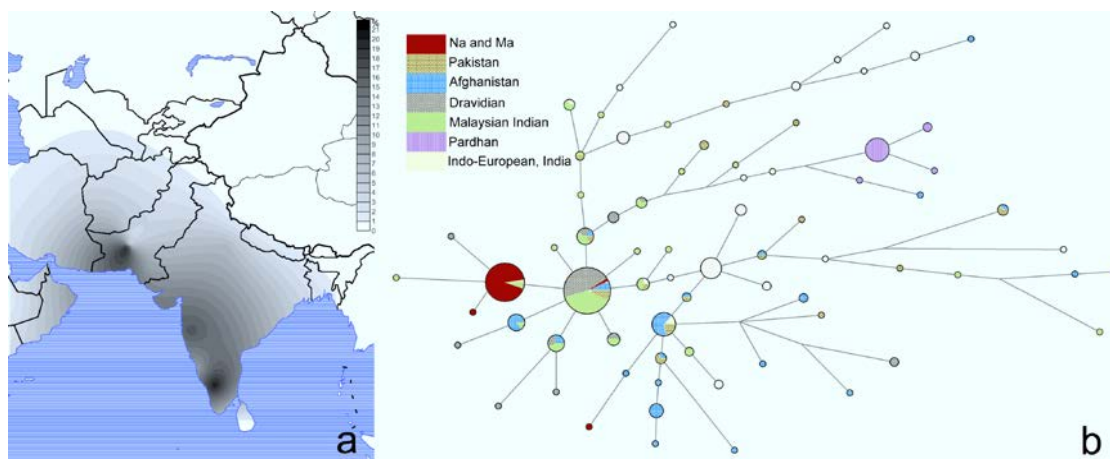

Figure 1. (a). Frequency distribution of haplogroup L1a-M76. The map was created by using the Kriging algorithm of the Surfer 9.11 package based on Google maps. (b) Median-joining network based on seven Y-STRs (DYS19, DYS389I, DYS389b, DYS390, DYS391, DYS392, and DYS393) of haplogroup L-M11 individuals. Reference population data on the Y chromosomes were retrieved from the literature[10-14].

## Acknowledgements

This work was supported by the National Natural Science Foundation of China (31071098, 91131002), National Excellent Youth Science Foundation of China (31222030), Shanghai Rising-Star Program (12QA1400300), Shanghai Commission of Education Research Innovation Key Project (11zz04), China Ministry of Education Major Project (311016) and Shanghai Professional Development Funding (2010001).